\documentclass[reprint,
nofootinbib,
 amsmath,amssymb,
 aps,
]{revtex4-2}

\usepackage{hyperref}
\usepackage[normalem]{ulem}
\usepackage{orcidlink}
\usepackage{graphicx}
\usepackage{dcolumn}
\usepackage{bm}
\usepackage{siunitx} 
\usepackage{graphicx}     
\usepackage{makecell}     
\usepackage{amsmath}      
\usepackage{xcolor}
\usepackage[compat=1.1.0]{tikz-feynman}
\usepackage{adjustbox}
\usepackage{floatrow}
\usepackage{multirow}
\usepackage{float}
\usepackage{cancel}
\usepackage{soul}
\usepackage{subfig}
\usepackage{color}
\usepackage{wrapfig}
\usepackage{caption}
\newcolumntype{C}[1]{>{\centering\arraybackslash}p{#1}}
\newfloatcommand{capbtabbox}{table}[][\FBwidth]

\usepackage{slashed}
\usepackage{pifont}

\usepackage[many]{tcolorbox}
\newtcolorbox{cross}{blank,breakable,parbox=false,
  overlay={\draw[black,line width=1pt] (interior.south west)--(interior.north east);
    \draw[black,line width=1pt] (interior.north west)--(interior.south east);}}
\usepackage{ulem,lipsum}
\newcommand\xoutpars[1]{\let\helpcmd\xout\parhelp#1\par\relax\relax}
\newcommand\soutpars[1]{\let\helpcmd\sout\parhelp#1\par\relax\relax}
\long\def\parhelp#1\par#2\relax{%
  \helpcmd{#1}\ifx\relax#2\else\par\parhelp#2\relax\fi%
}

\begin{document}

\preprint{APS/123-QED}

\title{Vector-like dark matter within an alternative left-right symmetric model}

\author{Yassine Bouzeraib~\orcidlink{0009-0004-7227-5641}}
\email{yassine.bouzeraib@univ-jijel.dz}
\affiliation{LPTh, Department of Physics, Faculty of Exact and Computer Sciences, University of Jijel, B. P. 98 Ouled Aissa, 18000 Jijel, Algeria.\\}
\author{Mohamed Sadek Zidi~\orcidlink{/0000-0003-4201-7655}}%
\email{mohamed.sadek.zidi@univ-jijel.dz}
\affiliation{LPTh, Department of Physics, Faculty of Exact and Computer Sciences, University of Jijel, B. P. 98 Ouled Aissa, 18000 Jijel, Algeria.}
\author{Geneviève Bélanger~\orcidlink{0000-0002-9621-4948}}
\email{genevieve.belanger@lapth.cnrs.fr}
\affiliation{Laboratoire d’Annecy de Physique Théorique, CNRS-USMB, 74940 Annecy, France}

\date{\today}

\begin{abstract}
We investigate an extension of the left-right symmetric model featuring  an additional non-abelian $SU(2)$ gauge symmetry. The particle content is augmented by one generation of vector-like leptons transforming under the fundamental representation of this new gauge group. We demonstrate that the neutral component of the vector-like lepton multiplet naturally provides a viable and stable dark matter candidate. Stability is ensured by imposing a discrete parity symmetry that forbids mixing between the vector-like leptons and the Standard Model leptons. As a consequence, the dark sector interacts with the visible sector exclusively through the vector portal (via s-channel processes)  and the vector-like lepton portal (via t-channel processes). In our analysis, we incorporate collider constraints on the mass of the first-generation extra charged gauge boson $W^{\prime\pm}$, while assuming that additional scalar states are decoupled from the relevant energy scale for simplicity. We identify the regions of parameter space consistent with the observed relic abundance,  collider bounds  on the charged partner
$E^{\pm}$, current direct detection limits from the LZ experiment and indirect detection constraints from Fermi-LAT.  We find viable dark matter with a mass at the TeV scale. We show the complementarity of direct  and indirect searches in probing the remaining parameter space of the model, in particular comparing the prospects of multi-ton direct detection experiments such as XLZD and of the CTA telescope.
\end{abstract}

\maketitle


\section{Introduction}
\label{int}
\noindent
The left-right symmetric model (LRSM)~\cite{Pati:1974yy,Mohapatra:1974gc,Senjanovic:1975rk,P,a,D,M} is a well-known extension of the Standard Model (SM) that has been developed over several decades, primarily to account for parity violation in weak interactions and to provide an explanation for neutrino masses through the seesaw mechanism~\cite{seesaw,seesaw1}.  Despite its theoretical appeal, the LRSM does not automatically provide a viable dark matter (DM) candidate. It is therefore natural to investigate whether the framework can be extended to address the dark matter problem.\\

\noindent
Many attempts have been made to incorporate a DM candidate within the LRSM framework. One possibility consists in lowering the mass of a right-handed neutrino (RHN) to the keV scale, thereby rendering it a long-lived warm DM candidate~\cite{Miha}. Alternatively,  imposing a discrete $Z_2$ symmetry on the scalar triplets, under which the left triplet $\Delta_L$ is  odd and the right triplet $\Delta_R$ is even, while setting the vacuum expectation value (vev) of the left triplet to zero. In the latter case, the neutral component  $\Delta_L^0$
becomes a viable DM candidate~\cite{Heeck}.	Unfortunately, the first approach is rather unnatural within the LRSM framework, while in the second approach $\Delta_L^0$ cannot reproduce the observed relic density due to its small annihilation cross section~\cite{Guo}.\\

\noindent
These limitations have motivated various extensions of the LRSM to address the DM problem. For example, the LRSM can be extended by adding a scalar singlet~\cite{Guo2} or a fermionic singlet~\cite{Patra} that plays the role of DM. Other possibilities include left-right fermion triplets and quintuplets forming a viable two-component DM scenario~\cite{Heeck}. Additional proposals involve stable fermionic or scalar multiplets whose stability is ensured purely by the gauge structure, without the need for an ad-hoc stabilizing symmetry~\cite{ Heeck2}, or models in which the gauge group is extended to $SU(3)_C\times SU(2)_L\times U(1)_L\times SU(2)_R\times U(1)_R$~\cite{Bhattacharyya:2022trp} containing a scalar bi-doublet and a singlet fermionic DM candidate. Models in which DM arises as a mixture of two or more multiplets have also been studied~\cite{Berlin:2016eem} as well as scenarios where the fermion sector is extended by an additional heavy right-handed copy, with the lightest heavy neutrino acting as DM~\cite{Dev:2016qeb}. Another possibility, which we explore in this work, is to extend the LRSM by introducing vector-like leptons (VLLs)~\cite{Bahrami}.\\

\noindent
In this paper, we extend the gauge structure of the LRSM by introducing an additional $SU(2)$ gauge symmetry, under which one generation of vector-like fermion (VLFs) doublet is charged. The model, first introduced in Ref.~\cite{Bouzeraib:2026nym}, aims to address several open questions, including the origin of parity violation, neutrino mass generation and the nature of dark matter. Here, we are primary interested in the possibility that the vector-like neutrino ($N$) could serve as a DM candidate. To ensure cosmological stability of $N$, we assume the existence of a new symmetry in the dark sector. However, imposing a discrete $Z_2$ symmetry -- commonly invoked in the literature to stabilize DM -- is not viable in this scenario due to the structure of the Yukawa interactions between the VLFs, the chiral fermions, and the scalar sector. To overcome this issue, we introduce a new parity symmetry that forbids mixing between the VLLs and the chiral leptons~\cite{Bahrami}. As a consequence,  the dark sector particles interact exclusively through the vector bosons portal or through the VLLs portal. The dark sector consists of the VLLs, namely  the DM candidate $N$ and its charged partner $E^{\pm}$. For simplicity,  we assume that the scalar sector is very heavy and effectively decoupled from the energy scale of interest. Moreover, we neglect the presence of vector-like quarks (VLQs), as their inclusion does not significantly affect the constraints on the extra gauge bosons derived in Ref.~\cite{Bouzeraib:2026nym}. The most stringent bounds arise from gauge boson decays into the second generation of heavy neutrinos (HNs), which are unaffected by the presence of VLQs.\\

\noindent
As in other models where RHNs couple to new gauge bosons~\cite{Belanger:2007dx}, the DM relic density falls within the narrow range measured by PLANCK~\cite{Planck:2018vyg} when the total mass of the initial particles in the annihilation (or co-annihilation) processes is close to the mass of the mediator, namely one of the neutral or charged gauge bosons of the model. Our goal in this paper is to assess the viability of this LRSM extension in light of current dark matter constraints. After imposing collider limits on the new gauge bosons ~\cite{Bouzeraib:2026nym}, we require that the predicted relic density does not exceed the upper bound determined by PLANCK observations. Additional constraints from LEP and LHC searches for new fermions, together with recent limits from direct detection (DD) experiments such as LUX-ZEPLIN (LZ)~\cite{LZ:2024zvo}, push the DM mass above the TeV scale. We also consider limits from dark matter indirect detection (ID) searches for  gamma rays from dwarf spheroidal galaxies by Fermi-LAT~\cite{Bonnivard:2015xpq, Alvarez:2020cmw}. However, these mainly constrain the low-mass region, which is already excluded by collider and cosmological bounds.\\

\noindent
Finally, we evaluate the prospects for probing this model in future experiments. In particular, we consider projected sensitivities from next-generation liquid xenon direct detection experiments such as XLZD/DARWIN~\cite{Aalbers:2022dzr}, and from indirect detection experiments such as CTA~\cite{CTA:2020qlo}, which are especially relevant for TeV-scale dark matter.\\

\noindent
The paper is organized as follows. The model is briefly reviewed in Sec.\ref{sec2}. Section \ref{DM_int} focuses on properties of the dark sector. Section \ref{Cold_const} includes constraints on the gauge bosons masses as well as other collider constraints. Section \ref{DM_obs} includes a discussion of the dark matter observables for a few benchmarks while the results of a general scan of the model defining the currently allowed parameter space as well as future probes is performed in Sec.\ref{GS}. Section \ref{conc} contains our conclusions.\\

\section{The model}
\label{sec2}
\noindent
We consider an extension of the LRSM which includes one generation of vector-like leptons. The VLLs belong to doublets of an extra gauge symmetry $SU(2)_V$. The vector-like nature of the VLLs requires that  both their chiral components belong to the same representation (the fundamental in this case). Consequently, their masses can be included directly in the Lagrangian without spoiling gauge invariance.  The left and right chiral components of SM fermions belong to  doublets under the fundamental representations of $SU(2)_L$ and $SU(2)_R$, respectively, as invoked  in the LRSM. Under the $SU(2)_V\times SU(2)_L\times SU(2)_R\times U(1)_{B-L}$ gauge group, the VLLs transform as,
 \begin{align}
L_{L,R}=\begin{pmatrix}
N \\
E
\end{pmatrix}_{L,R}\sim (2,1,1)_{-1}
 \end{align}
where the subscript "-1" denotes the $B-L$ quantum number.

\noindent
The gauge bosons fields and the gauge couplings of the model are denoted by,
 \begin{align}
  &SU(2)_V: W^i_{V}, g_{_V} &&
  &SU(2)_L: W^i_{L}, g_{_L} \nonumber\\
  &SU(2)_R: W^i_{R}, g_{_R} &&
  &U(1)_{B-L}: B, g^{\prime}
 \end{align}
where $i=1,2,3$ is the index of the gauge bosons components\footnote{In the mass basis, the gauge bosons include those of the standard model $\gamma, W^\pm,Z$, as well as four heavier gauge bosons $W^{\prime\pm}, Z^{\prime}, W^{\prime\prime\pm}$ and $Z^{\prime\prime}$.}.
\noindent
The left-right symmetry imposes the following conditions on the left and right-handed parts of the fermionic and gauge fields,
\begin{align}
  \Psi_L \longleftrightarrow \Psi_R && \vec W_{L\mu}\longleftrightarrow\vec W_{R\mu}
 \end{align}
which leads to
\begin{align}
 g_{_L}=g_{_R}=g
\end{align}
where $g$ is the SM weak coupling.\\

\noindent
 The scalar sector of the model contains a bi-doublet field $\Phi$ and left (right) triplets $\Delta_{_{L}}$ ($\Delta_{_{R}}$), which are introduced to spontaneously break the left/right symmetry of the model. These fields generate the chiral fermions interaction in the Yukawa sector and provide Majorana mass term for the neutrinos. To break the $SU(2)_V$ symmetry and allow the VLLs to mix with the chiral leptons, we need to introduce the two following self-dual bi-doublet scalar fields,
   \begin{align}
  \Phi_{_{VL}}=\begin{pmatrix}
  \phi^0_{_{1}} &  -\phi^{+}_{_{1}}\\
   \phi^{-}_{_{1}} & \phi^{0*}_{_{1}}\\
   \end{pmatrix}\sim(2,2,1)_{0}
   \nonumber\\
  \Phi_{_{VR}}=\begin{pmatrix}
  \phi^0_{_{2}} &  -\phi^{+}_{_{2}}\\
   \phi^{-}_{_{2}} & \phi^{0*}_{_{2}}\\
   \end{pmatrix}\sim(2,1,2)_{0}
    \end{align}

\noindent
We recall that the vev's of $\Delta_{_{R}}$ and $\Phi_{_{VR}}$ can be taken to be zero as shown in Ref.~\cite{Bouzeraib:2026nym}. The pattern of the symmetry breaking  of the model is schematized as follows:
\begin{align}
&SU(2)_V\times SU(2)_L\times SU(2)_R\times U(1)_{B-L}\xrightarrow[\langle\Phi_{_{VL}}\rangle=0]{\langle\Phi_{_{VR}}\rangle=u_{_{R}}} \nonumber\\ &SU(2)_L\times SU(2)_R\times U(1)_{B-L} \xrightarrow[\langle\Delta_L\rangle=0]{\langle\Delta_R\rangle=v_R}\nonumber\\
&SU(2)_L\times U(1)_{Y}\xrightarrow{\langle\Phi\rangle=\text{diag}(k_1,k_2)}U(1)_{EM}
\label{patern1}
 \end{align}
where the vev's $u_{_R}$ and $v_{R}$ are both at the TeV scale, and $k_1^2+k_2^2=v_{_{EW}}^2$, where $v_{_{EW}}$ is the SM vev. The bi-doublet $\Phi$ vev's can be parametrized as follows:
\begin{align}
 k_1=v_{_{EW}}\,c_\beta &&  k_2=v_{_{EW}}\,s_\beta
\end{align}
We assume $k_2\ll k_1$, following Ref.~\cite{Zhang:2007fn} \footnote{To ensure the known mass hierarchy for the ordinary top and bottom quarks, we must take $k_2\ll k_1$, see ref~\cite{Zhang:2007fn} for more details.}. Thus, the angle $\beta$ is small (i.e. $\sin(\beta)=s_{\beta}\sim 0$ and $\cos(\beta)=c_{\beta}\sim 1$).
The  symmetry breaking hierarchy ($u_{_R}>v_{R}$) imposes that the masses of the 2nd-generation extra-gauge bosons ($W^{\prime\prime}$ and $Z^{\prime\prime}$)  be larger than those of the 1st-generation ($W^{\prime}$ and $Z^{\prime}$).  At first order in $\epsilon_1$ and $\epsilon_2$, whith,
\begin{align}
 \epsilon_1=v_{_{EW}}^2/u_{_R}^2\ll1 && \epsilon_2=v_{_{EW}}^2/v_R^2\ll1
 \label{epsi_rel}
\end{align}
the gauge bosons mass squared are given by (cf.~Ref.~\cite{Bouzeraib:2026nym}):
\begin{align}
m^2_W\simeq&\frac{1}{4}g^2\,v^2_{_{EW}}\nonumber\\
m^2_{W^{\prime}}\simeq&\frac{g^2\,g_{_V}^2}{4(g^2+g_{_V}^2)}(v^2_{_{EW}}+2\,v^2_{R}) \nonumber\\
m^2_{W^{\prime\prime}}\simeq&\frac{1}{4(g^2+g_{_V}^2)}\Bigl(2(g^2+g_{_V}^2)^2\,u^2_{_{R}}+g^4(v^2_{_{EW}}+2\,v^2_{R})\Bigr)\nonumber\\
m^2_Z\simeq&\frac{1}{4}g^2\,\frac{v^2_{_{EW}}}{c^2_{W}}\nonumber\\
m^2_{Z^{\prime}}\simeq& \frac{1}{4}\Bigl(-\frac{e^2}{c^2_{W}}+\frac{g^2\,g_{_V}^2}{g^2+g_{_V}^2}\Bigr)v^2_{_{EW}}\nonumber\\
&-\frac{c^2_{W}\,g^4\,g_{_V}^4}{(g^2+g_{_V}^2)\bigl(e^2\,g^2+(e^2-c^2_{W}\,g^2)g_{_V}^2\bigr)}v^2_{R}\nonumber\\
m^2_{Z^{\prime\prime}}\simeq& \frac{g^2}{4\,(g^2+g_{_V}^2)}\,(4\,g_{_V}^2\,u^2_{_{R}}+g^2(v^2_{_{EW}}+4\,u^2_{_{R}}+4\,v^2_{R})
\end{align}
where $c_{W}$ and $e$ are the cosine of the Weinberg angle and the electromagnetic coupling, respectively. We recall that the gauge couplings are related by:
\begin{align}
g_L=g_R=g=\frac{e}{s_{W}} && g^{\prime}=\frac{g\,g_{_V}\,s_{W}}{\sqrt{c^2_{W}\,g_{_V}^2-s^2_{W}(g^2+g_{_V}^2)}}
\label{eq1}
\end{align}
The gauge coupling $g_{_V}$ is a free parameter, it is constrained from above using the perturbative unitarity condition $g_{_V}<\sqrt{4\,\pi}$. It is also bounded from below by requiring that $g^{\prime}$ in Eq.~(\ref{eq1}) be real, therefore,
\begin{align}
 g_{_V}>\frac{2\,\sqrt{\pi}\,s_{W}\,g}{\sqrt{4\,\pi\,(c^2_{W}-s^2_{W})-s^2_{W}\,g^2}}.
\end{align}
We emphasize that the last relation ensures that both  $g^{\prime}$ is real and the perturbative unitarity condition is satisfied.\\

\noindent
The mixing of the VLLs with the chiral leptons is achieved through the self-dual bi-doublet scalar fields. Thus, the Yukawa Lagrangian describing this mixing is given by:
\begin{align}
 {\cal{L}}_{int}=-\overline{L}_R \lambda^{\dagger}_l\,\Phi_{_{VL}}\,l_L-\overline{L}_L \lambda^{\prime}_l\,\Phi_{_{VR}}\,l_R+h.c
\end{align}
where $\lambda_{l}$ and $\lambda^{\prime}_{l}$ are $3\times 1$ ($1\times 3$) Yukawa coupling matrices.\\

\noindent
The scenario where $N$ plays the role of DM requires that it must be stable on cosmological times scales. This stability strongly suggests the existence of a new symmetry in the dark sector. There are many possibilities for symmetries that stabilize DM, for example the discrete $Z_2$ symmetry which is commonly used~\cite{LopezHonorez:2006gr,LopezHonorez:2010eeh,LopezHonorez:2010tb,Batell:2010bp,Hirsch:2010ru,Lavoura:2012cv,Earl:2013fpa,Ma:2014eka,Baek:2014kna,Lamprea:2016egz,Belanger:2022gqc}. Obviously, this possibility cannot be used in our case because  this symmetry imposes that both the VLLs and the self-dual bi-doublets must be odd-fields (i.e. $L_{L,R}\rightarrow -L_{L,R}$, $\Phi_{VL(R)}\rightarrow -\Phi_{VL(R)}$) under $Z_2$, while all other fields should be even. $\Phi_{VL}$ could be an odd-field when their vev is zero, $\langle\Phi_{VL}\rangle=0$. In contrast, the vev of $\Phi_{VR}$ should be at TeV scale hence it could not be an odd-field under $Z_2$. To circumvent this problem one can  include  a new parity symmetry which disallows mixing between the VLLs and the ordinary leptons~\cite{Bahrami}, under this symmetry only the VLLs are odd-fields while all other fields  are even, such that $\lambda_{l}=\lambda^{\prime}_{l}=0$. The latter symmetry implies independent free masses for the vector-like neutrino $N$ and their charged partner $E^{\pm}$. As will be seen in the following sections, their masses will be constrained by phenomenological requirements, mainly by imposing the  DM relic density constraint as well as the constraints from the DD and ID approaches. In this scenario neutrinos  get their masses through the see-saw mechanism as invoked usually within the LRSM since there is no mixing with the VLL~\cite{seesaw,seesaw1,a,P}.\\

\noindent
In the following sections, we will examine various experimental constraints on the relevant parameter space for the DM study, for this it is more convenient to use the physical parameters as free parameters. The free parameters include the coupling $g_{_V}$,  the vev of $\langle\Phi_{VR}\rangle$ ($u_{_R}$), as well as the the mass of one extra gauge boson, $m_{W^{\prime}}$. In the following, we fix $u_R=v_R+100$~GeV, thus the masses of new gauge bosons can all be rewritten in terms of 
$g_{_V}$ and $m_{W^\prime}$. In addition, we fix the mass of the HNs to be $m_{W^\prime}/2$ since this relation is used to obtain the collider limits on the heavy gauge bosons, we keep the same feature  for the rest of the analysis. For the sake of simplicity, all the masses of the extra scalar bosons are considered to decouple from the scale of interest. Finally, in the dark sector, the two parameters relevant for computing DM observables are  the mass of the DM candidate $m_N$ and its mass  splitting  with  the charged partner $E^{\pm}$  which is denoted by $\Delta_m$.

\section{Dark sector interactions}
\label{DM_int}
\begin{figure}
\includegraphics[width=9cm,height=2cm]{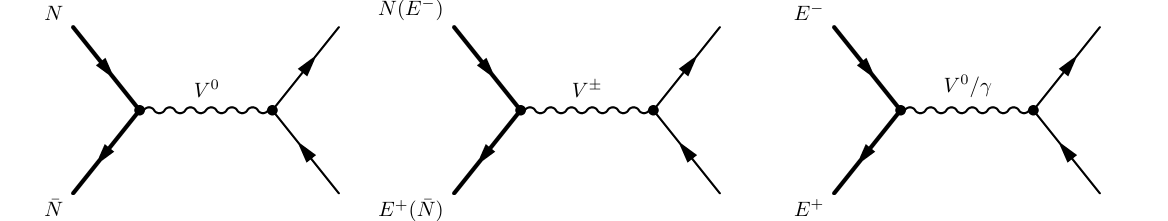}
\includegraphics[width=9cm,height=2cm]{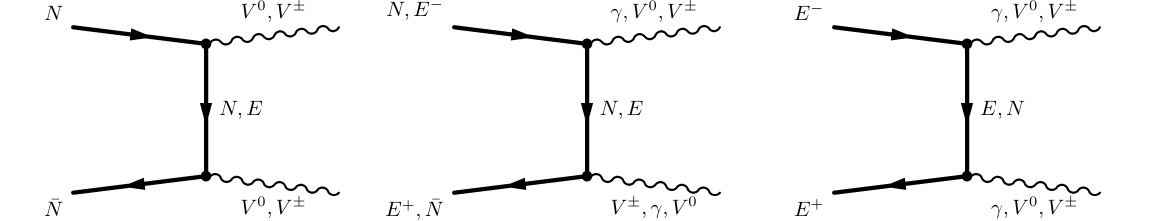}
\caption{\small Feynman diagrams for the annihilation and co-annihilation processes via the vector boson/vector-like lepton portals. In addition to fermions in the final state, the s-channel processes (up diagrams) can involve gauge bosons and/or the Higgs in, see eq.~(\ref{final_states}).}
\captionsetup{singlelinecheck=false, justification=raggedright}
\label{FeynD}
\end{figure}

\noindent
In the following, we discuss the interactions among  the dark sector particles and their relevance for DM annihilation and co-annihilation processes. There are two types of channels for which the annihilation and the co-annihilation of dark particles can proceed, first the s-channel processes mediated exclusively by the vector bosons, see Fig.\ref{FeynD} (upper), second the t-channel annihilation processes into vector bosons mediated by the VLLs of the dark sector, Fig.\ref{FeynD} (lower). Technically, any  combination of particles that interact with one of the gauge bosons (i.e. the fermions\footnote{Including the SM fermions ($f$) and the heavy neutrinos (HNs).}, the gauge bosons\footnote{Including the photon $\gamma$, the neutral gauge bosons $V^{0}=Z, Z^{\prime}, Z^{\prime\prime}$ and the charged gauge bosons $V^{\pm}=W^{\pm}, W^{\prime\pm}, W^{\prime\prime\pm}$.} and the Higgs boson\footnote{ Since the scalar sector is decoupled from the energy scale of interest for the sake of simplicity, the Higgs boson ($h$) is the only scalar that could appear in the final state.}) could be present in the final states of the s-channel processes  provided they are kinematically allowed. On the other hand, the final sates for the t-channel can be exclusively the gauge bosons.
This is a consequence of the symmetry imposed for stabilizing the DM particle which disallow  interactions of the VLLs through Yukawa couplings.

\begin{table*}[htbp]
\begin{tabular}{|c|c|c|c|}
\hline
\textbf{Vertex} & \textbf{Coupling expression} &
\textbf{Vertex} & \textbf{Coupling expression} \\
\hline
$N\overline{N}Z$&
$\frac{i\,g_{_V}}{8\,s_{\theta_w}}
\left(c_2^4\,s_1\,s_2\,\epsilon_2-2\,(c_2^2\,s_2^3 + c_1^2\,s_2^5)\,s_1^3\,\epsilon_1
\right)\gamma^{\mu}$&
$\overline{N} E^{-} W^{-}$&
$\frac{i\,g_{_V}}{\sqrt{2}}\, \cot(\theta_w)\, c_2\, s_1\, c_{\beta}\, s_{\beta}\, \epsilon_2\, \gamma^{\mu}$\\
& $+ \frac{i\,g^{\prime}}{8\,s_{\theta_w}}
\left(c_2^3\,s_2^3\,\epsilon_2-2\,c_2\,s_1^4\,s_2^4\,\epsilon_1\right)\gamma^{\mu}$&
$N E^{+} W^{+}$& \\
\hline 
$N\overline{N}Z^{\prime}$ &
$\displaystyle
\frac{i\,g_{_V}}{2}\,c_2\,s_1\,\gamma^{\mu}
+\frac{i\,g^{\prime}}{2}\,s_2\,\gamma^{\mu}$&
$\overline{N} E^{-} W^{\prime-}$&
$\displaystyle\frac{i\,g_{_V}}{\sqrt{2}}\,s_1\,\gamma^{\mu}$\\
& & $N E^{+} W^{\prime+}$ & \\
\hline 
$N\overline{N}Z^{\prime\prime}$ &
$\displaystyle\frac{i\,g_{_V}}{2}\,c_1\,\gamma^{\mu}$&
$\overline{N} E^{-} W^{\prime\prime-}$&
$\displaystyle\frac{i\,g_{_V}}{\sqrt{2}}\,c_1\,\gamma^{\mu}$\\
& & $N E^{+} W^{\prime\prime+}$ & \\
\hline 
$E^{+}E^{-}\gamma$&
$\displaystyle-i\,g\,s_{\theta_w}\,\gamma^{\mu}$&
$\displaystyle E^{+} E^{-} Z^{\prime}$&
$\displaystyle -\frac{i\,g_{_V}}{2}\,c_2\,s_1\,\gamma^{\mu}+\frac{i\,g^{\prime}}{2}\,s_2\,\gamma^{\mu}$ \\
\hline 
$E^{+} E^{-} Z$ &
$\frac{i\,g_{_V}}{8\,s_{\theta_w}}
\left(-c_2^4\,s_1\,s_2\,\epsilon_2+2\,(c_2^2\,s_2^3 + c_1^2\,s_2^5)\,s_1^3\,\epsilon_1\right)\gamma^{\mu}$&
$\displaystyle E^{+} E^{-} Z^{\prime\prime}$&
$\displaystyle -\frac{i\,g_{_V}}{2}\,c_1\,\gamma^{\mu}$\\
& $+ \frac{i\,g^{\prime}}{8\,s_{\theta_w}}
\left(c_2^3\,s_2^3\,\epsilon_2-2\,c_2\,s_1^4\,s_2^4\,\epsilon_1+4\,c_2\,s_{\theta_w}^2\right)\gamma^{\mu}$ & & \\
\hline 
\end{tabular}
\caption{\small Couplings relevant for annihilation and co-annihilation processes (with $s_i^2=1-c_i^2$ for $i=1,2$).}
\label{tab:NN_NE}
\end{table*}

\noindent
Although any combinations of the gauge bosons that satisfy the conservation of electric charge,  can be produced  in the final state for the co-annihilation processes, in our model the   $Z^{\prime\prime}/W^{\prime\prime\pm}$  are  kinematically forbidden. Moreover, one should mention that there is no interaction between the DM particle and the photon (i.e. in lower middle diagram of Fig.\ref{FeynD}). All possible final states for the s-channel diagrams (i.e. the upper diagrams of Fig.\ref{FeynD}) are listed here:
\begin{align}
&N\overline{N} (E^{+}E^{-})\to V^{0} \;\to\;
\begin{cases}
q\bar{q},\;l_i^{+}l_i^{-}\\
\nu_i \nu_i,\;\nu_i N_i,\;N_i N_i\\
W^+W^-,\;W^{\pm} W^{\prime\mp}\\
Z h,\;Z^{\prime}h
\end{cases} 
\nonumber\\
&N(\overline{N}) E^{\pm}\to V^{\pm} \;\to\;
\begin{cases}
q\bar{q}'\\
\nu_i l_i^{\pm},\;N_i l_i^{\pm}\\
\gamma W^{\pm}\\
Z W^{\pm},\;Z W^{\prime\pm},\;Z^{\prime} W^{\pm}\\
W^{\pm} h,\;W^{\prime\pm} h
\end{cases}
\nonumber\\
&\begin{cases}
N\overline{N} (E^{+}E^{-})\to Z/Z^{\prime}\;\to\;W^{\prime\pm} W^{\prime\mp} \\
N(\overline{N}) E^{\pm}\to W^{\pm}/W^{\prime\pm} \;\to\;\gamma W^{\prime\pm},Z^{\prime} W^{\prime\pm}
\end{cases}
\nonumber\\
&E^{+}E^{-}\to \gamma\;\to\;
\begin{cases}
q\bar{q},\;l_i^{+}l_i^{-}\\
W^+W^-,\;W^{\prime\pm} W^{\prime\mp}\\
W^{\pm} W^{\prime\mp}\\
\end{cases}
\label{final_states}
\end{align}
where $q$ represent the quarks and $i$  the generation index for the leptons (charged leptons ($l^{\pm}_i$), Majorana neutrinos ($\nu_i$) and HNs ($N_i$)), $V^0(V^\pm)$ stands for the neutral (charged) gauge bosons.\\

\noindent
The couplings for the  interaction of the VLLs with the gauge bosons are summarized in Table~\ref{tab:NN_NE}. They depend on the angles used for the diagonalization of  the mass matrices of the gauge sector in Ref.~\cite{Bouzeraib:2026nym}, namely:
\begin{align}
 c_1 &= \frac{g_{_V}}{\sqrt{g^2+g^2_{_V}}} &
 c_2 &= \frac{g\,g_{_V}}{\sqrt{g^{\prime 2}\left(g^2+g_{_V}^2\right)+g^2\,g_{_V}^2}}.
\end{align}
Clearly, the interactions via $Z$ and $W^{\pm}$ are suppressed since they are proportional to the suppressed parameters $\epsilon_{1,2}$ and $s_\beta$ defined in Eq.~\ref{epsi_rel}. Moreover, the $N\overline{N}Z$ coupling is very strongly suppressed  because of the multiple powers of small mixing angles entering its definition. This interaction do not contribute significantly to the dark matter formation as we will see. Thus, the efficient annihilation can be realized exclusively via $Z^{\prime}$ and $Z^{\prime\prime}$. On the other hand,  efficient  co-annihilation can proceed  via the three charged gauge bosons ($V^{\pm}$) although the interaction with $W^{\pm}$ is somewhat suppressed in comparison with the others.

\section{Colliders constraints}
\label{Cold_const}

\subsection{LHC constraints on the bosons masses}
\label{LHC_const}
\noindent
Current constraints from the LHC on new heavy gauge bosons were derived in Ref.~\cite{Bouzeraib:2026nym} and  Table $1$ shows the experimental limits on $m_{W^{\prime}}$ (consequently $m_{Z^{\prime}}$). The most restricting limits arise from the $N_2\mu^{\pm}$ channel (i.e. $pp\to W^{\prime\pm}\to N_2\mu^{\pm} (N_2\to \mu^{\pm}  q \bar q^{\prime})$). These results could be exploited for getting an indirect  lower limit on the 2nd-generation extra-gauge bosons ($W^{\prime\prime\pm}$ and $Z^{\prime\prime}$) masses. Since the pattern of symmetry breaking is chosen such that ($u_{_R}>v_{R}$), we obtain the following mass hierarchies  ($m_{W^{\prime\prime}}$ and $m_{Z^{\prime\prime}}$) $>$ ($m_{W^{\prime}}$ and $m_{Z^{\prime}}$). Moreover by fixing $u_{_R}=v_{R}+100$ GeV for covering as many as possible benchmark points during the analysis, one is able to derive a lower bound on all the gauge boson masses from 
 the experimental lower limit on $m_{W^{\prime}}$.  The results are represented in Table~\ref{tab:mass_limits}.
 
\begin{table}[!htbp]
\begin{tabular}{|l|c|c|c|c|}
\hline
$g_{_V}$ & $m_{W^{\prime}}$ [TeV] & $m_{Z^{\prime}}$ [TeV] & $m_{W^{\prime\prime}}$ [TeV] & $m_{Z^{\prime\prime}}$ [TeV] \\
\hline
$g$ & $3.0$ & $6.5$ & $6.8$ & $7.4$ \\
$1$   & $3.4$ & $6.3$ & $7.8$ & $8.1$ \\
$2$   & $3.7$ & $6.4$ & $12.7$ & $12.8$ \\
$3$   & $3.8$ & $6.4$ & $18.2$ & $18.3$ \\
\hline
\end{tabular}
\caption{Lower mass limits on $W^{\prime\pm}$, $Z^{\prime}$, $W^{\prime\prime\pm}$, and $Z^{\prime\prime}$ for different benchmark values of $g_{_V}$, based on CMS data on $W^{\prime}$ production and decay via the $N_2 \mu$ channel \cite{CMS:2021dzb}. The condition $u_{_R} = v_{_R} + 100~\mathrm{GeV}$ is satisfied.}
\label{tab:mass_limits}
\end{table}

\noindent
The results in this table were derived for a benchmark point where the mass of the three generations of the HNs are $m_{N_i}=m_{W^{\prime}}/2$, thus we will keep this assumption for the rest of the analysis. Furthermore, all the masses of the extra bosons from the scalar sector were fixed at a scale of several tens of TeV's.

\subsection{LEP constraints on $E^{\pm}$ mass}
\noindent
Two types of constraints from LEP on charged leptons are relevant. First, the precise measurement of the $Z$ boson partial decay width at LEP1, $\Gamma(Z\to E^-E^+)< 2$ MeV~\cite{ALEPH:2001hzl} rules out the region where  the $Z$ boson could decay into $E^-E^+$ that is  when  $m_E<m_Z/2$.  Second, when $m_Z/2< m_E \lesssim 104$~GeV, reinterpreting the LEP 2~\cite{ALEPH:2002gap} limits on chargino mass allow us to exclude all the points in this region. We can  apply directly these results  because the production cross section is dominated by $\gamma$ exchange which is a model independent process. We apply these two limits from LEP when performing a general scan over the parameter space in section~\ref{GS}, we expect strong constraints on the new charged VLL lighter than 104 GeV.

\subsection{LHC constraints on $E^{\pm}$ mass}
\noindent
After imposing the relic density constraint we will see that all the point allowed at the electroweak scale receive an important contribution from co-annihilation channels and therefore predict  
a small mass splitting $\Delta_m$  between the charged lepton and DM (see Fig.~\ref{fig:RM}). Thus, constraints on the mass of the charged VLL $E^{\pm}$  from searches for long-lived charged particles  at the LHC \cite{ATLAS:2019gqq,CMS:2013czn,CMS:2015lsu} will strongly impact the possibility of having  DM below the TeV scale. We used {\tt SModelS}~\cite{Altakach:2024jwk} as implemented in {\tt micrOMEGAs}~\cite{Alguero:2023zol} to apply the LHC constraints on VLL's. We found that only a few points with a DM mass below the TeV scale evade the LHC constraints, in all cases the DM mass is around $130$ GeV.  Note however that as we will see in section~\ref{GS}, these same points will  be excluded by DM direct detection.

\section{DM observables: fixed gauge boson masses}
\label{DM_obs}

\begin{figure*}[htbp]
  \centering
  {\includegraphics[width=8cm,height=4.5cm]{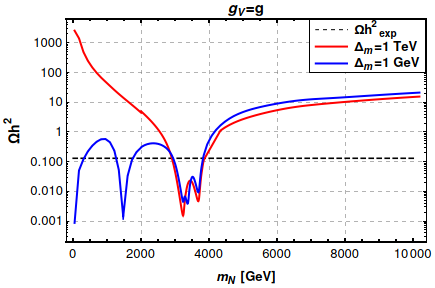}}
  {\includegraphics[width=8cm,height=4.5cm]{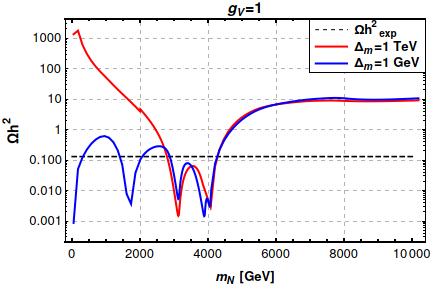}}
   {\includegraphics[width=8cm,height=4.5cm]{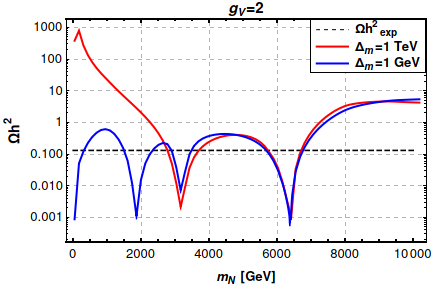}}
  {\includegraphics[width=8cm,height=4.5cm]{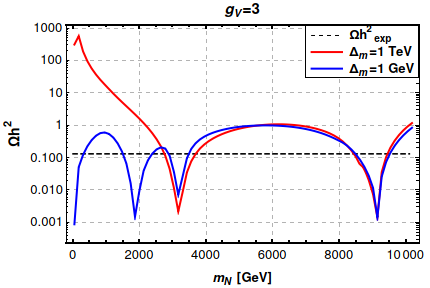}}
  \caption{$\Omega h^2$ variation w.r.t $m_N$ for several benchmark values of the gauge coupling $g_{_V}$.}
  \label{fig:DM}
\end{figure*}

\noindent
In this section, we calculate some DM observables for fixed gauge boson masses. We recall that {\tt FeynRules} package~\cite{Alloul:2013bka} is used to generate the model in {\tt CalcHEP} \cite{Belyaev:2012qa} format. Through this article, we use {\tt micrOMEGAs}~\cite{Alguero:2023zol} to calculate DM observables.
\subsection{Relic density}
\label{DM_RD}

\noindent
The relic density $\Omega h^2$ of $N$ provides one of the most important experimental constraints on the model. We impose the condition  that the total relic density falls within the observed range determined by the PLANCK collaboration \cite{Planck:2018vyg},
\begin{align}
\Omega h^2=0.1184\pm 0.0012
\end{align}
We allow for a theoretical uncertainty of order $10\%$ ($\Omega h^2=[0.11,0.13]$) as estimated in several studies \cite{Banerjee:2019luv, Banerjee:2021oxc, Belanger:2022esk}. 
Fixing the gauge boson masses at their lower limit as listed in Table \ref{tab:mass_limits}, we display in Fig.\ref{fig:DM} the predicted value of $\Omega h^2$ as a function of the DM mass for different choices of the coupling $g_{_V}$. We can see from the different panels that DM is generally overabundant unless the mass of the DM  lies around half of one of the vector boson mass. We choose two scenarios: the  first features a large mass splitting between the DM and its charged partner $E^{\pm}$,  $\Delta_m=1$ TeV. For this case the dominant channels are the annihilation ones  mediated by the neutral gauge bosons. We notice that the relic density drops sharply   when the mass of the DM becomes near $m_{Z^{\prime}}/2$ and $m_{Z^{\prime\prime}}/2$. However,  the strongly suppressed coupling of two DM  to the $Z$ boson, does not allow to attain  $\Omega h^2\approx 0.12$ when the DM mass is  $m_{Z}/2$. In the second scenario, we take the mass splitting  to be $\Delta_m=1$ GeV. Thus,  co-annihilation effects becomes notable around half of the charged gauge bosons masses (see the blue lines in Fig.\ref{fig:DM}). The interaction of the DM particle and their charged partner via $W^{\pm}$ is suppressed as shown in Table~\ref{tab:NN_NE}, but still lead to a correct value of $\Omega h^2$ at the threshold of the process.

\subsection{Direct detection}
\label{DM_SI}
\begin{figure}
\includegraphics[width=5cm,height=2.0cm]{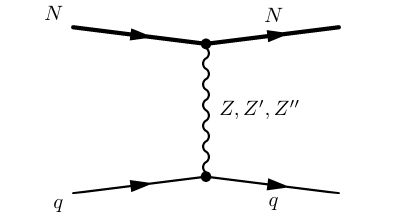}
\caption{\small Feynman diagrams for the DM elastic scattering off nucleons.}
\label{SI:cs}
\end{figure}

\noindent
As mentioned above, the symmetry that enforces the Yukawa couplings of the VLLs-leptons to be zero with the aim of stabilizing the DM particle, entails that the DM candidate can interact with nucleons exclusively through the t-channel diagrams mediated by the neutral gauge bosons as shown in Fig.\ref{SI:cs}. For the same reason, there is no interaction of the DM with any scalar particles. Moreover, the structure of the model, where the VLLs belong to another gauge group  $SU(2)_V$ and the bi-doublet $\Phi$ -- from which the Higgs particle originates -- is chargeless under this group, entails that there is no interaction via the ordinary Higgs.

\begin{figure*}[htbp]
  {\includegraphics[width=8cm,height=5cm]{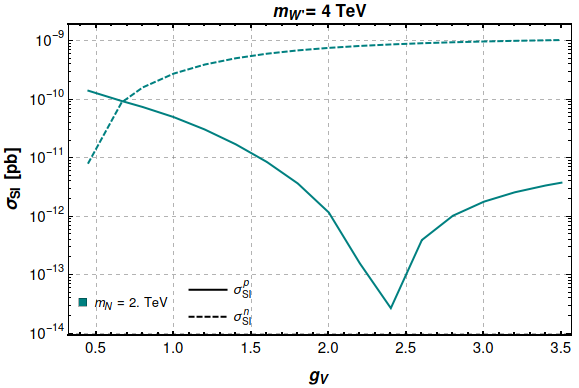}}
  {\includegraphics[width=8cm,height=5cm]{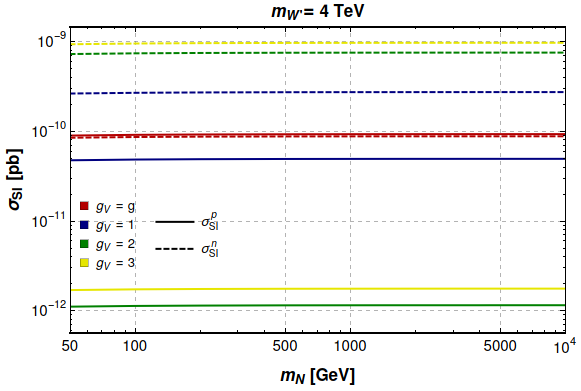}}
  \caption{\small (Left panel) $\sigma_{SI}$ off proton and neutron variation w.r.t $g_{_V}$ for a benchmark value of $m_{N}=m_{W^{\prime}}/2=2$ TeV. (Right panel) $\sigma_{SI}$ off proton and neutron variation w.r.t $m_N$ for a benchmark value of $m_{W^{\prime}}=4$ TeV and several values of $g_{_V}$.}
  \label{fig:gv:SI}
\end{figure*}

\vspace{0.15cm}
\noindent
DM interacts  with nucleons through spin-independent (SI) interactions and the elastic scattering cross sections differ for  neutrons ($\sigma_{SI}^{n}$) and protons ($\sigma_{SI}^{p}$).
The behavior of the cross sections is determined by the relative contribution of the $Z,Z'$ and $Z''$ exchange and of their interference. The $Z,Z'$ exchange diagram leads to a much larger cross section for neutrons than protons while the $Z''$ exchange gives similar contributions. The value of the gauge coupling $g_{_V}$ has a significant  impact on  both cross sections  as can be seen in the left panel of Fig.\ref{fig:gv:SI} .  At small values of $g_{_V}$, the $Z'$ exchange dominates and destructive interference with the $Z$ contribution suppresses $\sigma_{SI}^{n}$   such that $\sigma_{SI}^{p}$ is larger than $\sigma_{SI}^{n}$. As $g_V$ increases, $\sigma_{SI}^{p}$ decreases rapidly, mainly because the $Z'$ contribution decreases, thus for $g_V\approx 0.7$ one gets an identical cross section for protons and neutrons. 
For larger values of $g_V$, the contribution of $Z^{\prime}$ to $\sigma_{SI}^{n}$ becomes dominant and it increases steadily with $g_{_V}$ while the $Z/Z^{\prime\prime}$ exchange gives the subdominant contribution. For $\sigma_{SI}^{p}$ the behavior is quite different, for $g_V \approx 0.8$, there is a strong suppression of the $Z$ contribution while for $g_V \approx 1.6$ it is the $Z'$ contribution that is strongly suppressed. Taking into account all interference effects, the minimum of the cross section is found around $g_V=2.4$, for larger values of the coupling, all three gauge bosons  contribute, nevertheless  $\sigma_{SI}^{p}$ is two orders of magnitude smaller  than $\sigma_{SI}^{n}$.
These results were obtained for $m_N= 2$ TeV, note however that the same feature occur for other DM masses as the SI cross sections are basically independent of $m_N$ as can be seen in Fig.~\ref{fig:gv:SI} (right). 

\vspace{0.15cm}
\noindent
Since the neutrons or protons contributions are not necessarily equal, measuring the nuclear recoil energy from the elastic scattering off nucleus would provide appropriate constraints on parameter space of the model in comparing the results from the DD experiments. The current most restricted limits are coming from the LZ experiment~\cite{LZ:2024zvo} and the future projections from   XLZD~\cite{Aalbers:2022dzr}. Thus, computing the normalized  to one nucleon cross section for a point-like Xenon nucleus would be more convenient in aim to directly compared with the experimental limits \cite{Belanger:2011rs},
\begin{align}
\sigma_{SI}^{Xe}=\frac{4\mu_N^2}{\pi}\frac{\left(Z\,f_p+\left(A-Z\right)f_n\right)^2}{A^2}
\end{align}
 where $\mu_N=\frac{m_N\,m_n}{m_N+m_n}$ is the reduced mass of the DM with the nucleon ($n$), $Z$ $(A-Z)$ is the number of protons (neutrons) in Xenon nucleus and $f_{p,n}$ are the couplings to protons and neutrons.\\
 
\begin{figure*}[htbp]
  {\includegraphics[width=8cm,height=4.5cm]{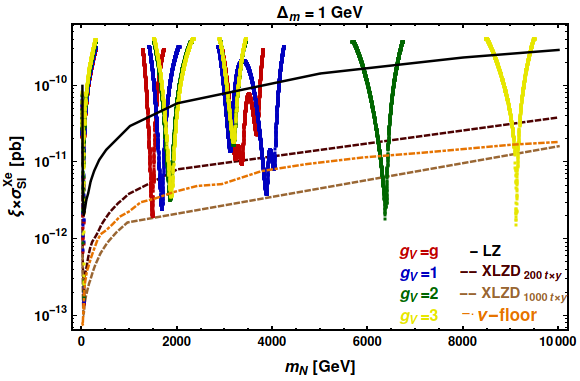}}
  {\includegraphics[width=8cm,height=4.5cm]{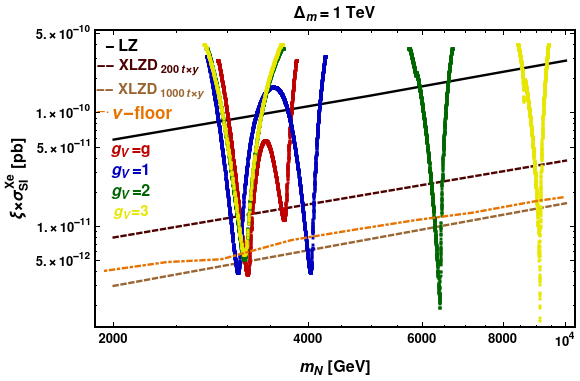}}
  \caption{$\xi\times\sigma_{SI}^{Xe}$ w.r.t $m_N$ for tow scenarios of mass splitting $\Delta_m=1$ GeV (TeV) and several benchmark values of $g_V$. The thick black line represent the current limits of LZ experiment \cite{LZ:2024zvo}, the brown dashed lines represent the future projection limits of XLZD for exposures $200t\times y$ and $1000t\times y$ \cite{Aalbers:2022dzr}, where the orange dash-dotted one represents the neutrino floor \cite{Billard:2013qya}. The inset in the left panel highlight the low mass  region of DM for $\Delta_m=1$ GeV  scenario.}
  \label{fig:Xe:SI}
\end{figure*}

\noindent
We fixed first the gauge bosons masses at their lower limits as provided in Table~\ref{tab:mass_limits} and fixed the masses of the HNs to be half of $m_{W^{\prime}}$ for each benchmark value of $g_{_V}$. Then using {\tt micrOMEGAs}, we made a random scan over $m_N$ where we imposed that the relic density constraints should be satisfied, weather our candidate explain all the DM in the universe or part of it. The results are shown in Fig.~\ref{fig:Xe:SI} where we plot the rescaled SI cross section off Xenon $\xi\times\sigma_{SI}^{Xe}$ w.r.t $m_N$, where $\xi$ is the fraction of the predicted value of the relic density over the observed value. Because of the contributions from both the annihilation and co-annihilation processes to the relic density,  the $\Delta_m=1$ GeV scenario (left panel) drops sharply when $m_N$ is around half of the neutral and charged gauge bosons masses. The valid points at the GeV scale appears since the relic density condition is satisfied because of the co-annihilation channels contribution that are mediated by the $W^{\pm}$ boson, and disappear as expected regarding to the other scenario when $\Delta_m=1$ TeV (right panel) where the relic density condition is not satisfied (see Fig.~\ref{fig:DM}).

The plots indicate that only the region near a gauge boson resonance is allowed  by current LZ limits, and that most of these regions are within  the reach of the future XLZD projections. Note however that  some of the points can not be probed since they fall below the neutrino floor~\cite{Billard:2013qya}. In Fig.~\ref{fig:Xe:SI} the neutral gauge bosons masses were fixed at their lower bounds, we expect that the SI scattering cross section will decrease for heavier masses (see Fig.~\ref{fig:SI_MG_Scan}) thus allowing a wider range of allowed points. In the next section we will explore the full parameter space of the model. In this general scan, we will apply also collider constraints, namely the LEP~1~\cite{ALEPH:2001hzl} and LEP~2~\cite{ALEPH:2002gap} limits, in addition to the LHC searches for long-lived charged particles~\cite{ATLAS:2019gqq,CMS:2013czn,CMS:2015lsu} embedded in {\tt SModelS}~\cite{Altakach:2024jwk}. As a result  the viable points at the electroweak scale will be strongly suppressed.  Moreover we will show that  the TeV scale DM can be constrained from indirect searches, in particular from the limits on DM annihilation into photons from CTA~\cite{CTA:2020qlo}.

\section{General scan}
\label{GS}
\noindent
The four free parameters for our study have been discussed in section~\ref{sec2}. The gauge coupling $g_{_V}$ is fixed to four benchmarks values, see Table~\ref{Parm:tab}. The remaining three parameters are varied randomly, taking into consideration the collider constraints on the gauge bosons masses from Table~\ref{tab:mass_limits}. The range considered is represented in Table~\ref{Parm:tab},

\begin{table}[h!]
\begin{center}
\begin{tabular}{|c|c|}
\hline
$g_{_V}$ &  $m_{W^{\prime}}$ \\
\hline
$g$ &  $3-10$ TeV  \\
\hline
$1$ &  $3.4-10$ TeV  \\
\hline
$2$ &  $3.7-10$ TeV \\
\hline
$3$ & $3.8-10$ TeV \\
\hline
\end{tabular}
\begin{tabular}{|c|c|}
\hline
$m_N$ &  $10$ GeV $-$ $10$ TeV \\
\hline
$\Delta_m$ &  $1$ MeV $-$ $1$ TeV \\
\hline
\end{tabular}
\caption{\small The range of the four free parameters used in the scan.}
\label{Parm:tab}
\end{center}
\end{table}

\subsection{Relic density}
\begin{figure*}[htbp]
  \centering
  {\includegraphics[width=8cm,height=4.25cm]{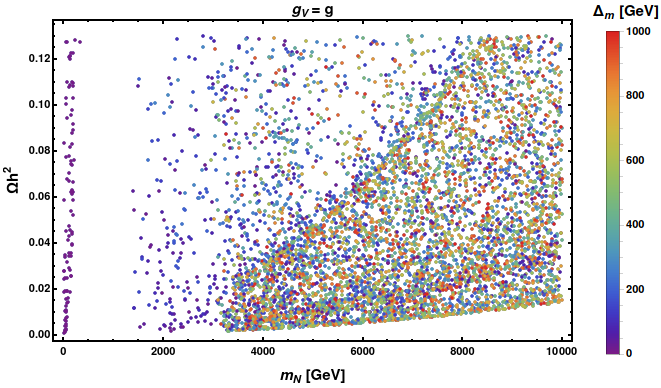}}
  {\includegraphics[width=8cm,height=4.25cm]{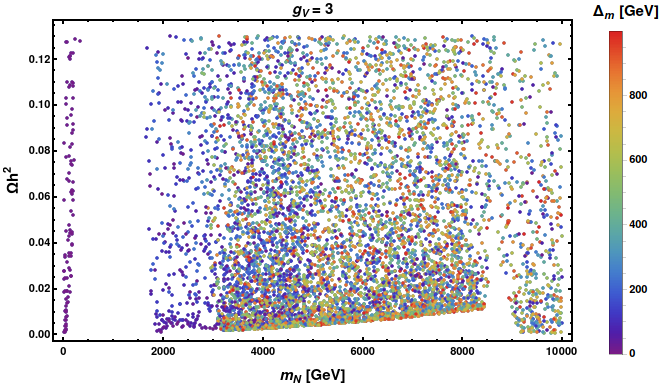}}
  \caption{$\Omega h^2$ variation w.r.t $m_N$ for two benchmark values of the gauge coupling $g_{_V}$, with the color palette indicating the splitting mass $\Delta_m$.}
  \label{fig:RM}
\end{figure*}

\noindent
For this general  scan, we impose only an upper bound on the value of the relic density, $\Omega h^2<0.13$. This value includes  a $10\%$ theoretical uncertainty  as in sub.sec~\ref{DM_RD}. As discussed in the previous section, DM (co-)annihilation is most efficient when the gauge boson exchanged in s-channel is near resonance. Since the masses of the new gauge bosons are above $3$ TeV, the upper bound on the relic density requires in general a dark matter mass above $1.5$ TeV.  Fig.~\ref{fig:RM} shows the predicted value of the relic density as a function of the DM mass for two different values of the gauge coupling $g_{_V}$. The mass splitting  $\Delta_m$ is indicated by the color palette. Fig.~\ref{fig:RM} also shows that a few points are found for a DM mass below $200$ GeV, these are associated with  a very small mass splitting between the dark matter and the charged VLL and correspond to co-annihliation channels mediated by $W^{\pm}$ exchange. Note that at this point we have not imposed the collider constraints on charged leptons mentioned in section \ref{Cold_const}. Such constraints will exclude most of the low mass region as will be discussed next.

\subsection{Direct detection and collider searches}
\begin{figure*}[htbp]
  {\includegraphics[width=8cm,height=4cm]{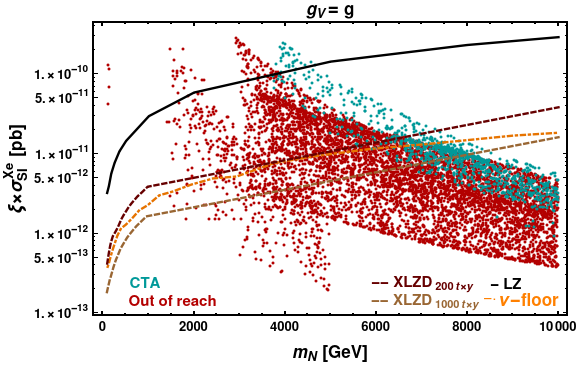}}
   {\includegraphics[width=8cm,height=4cm]{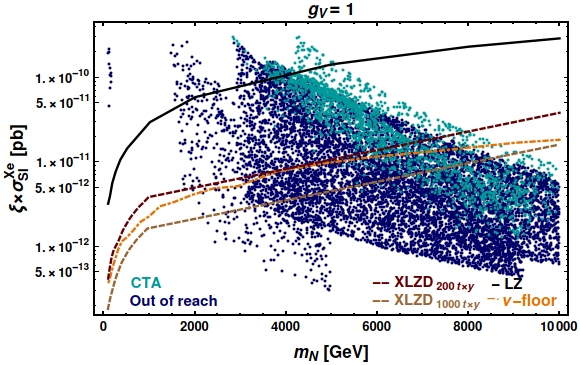}}
    {\includegraphics[width=8cm,height=4cm]{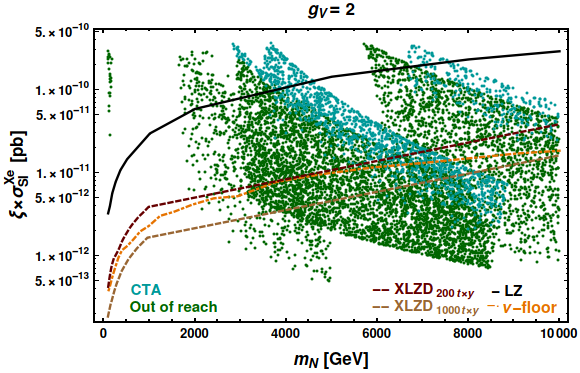}}
     {\includegraphics[width=8cm,height=4cm]{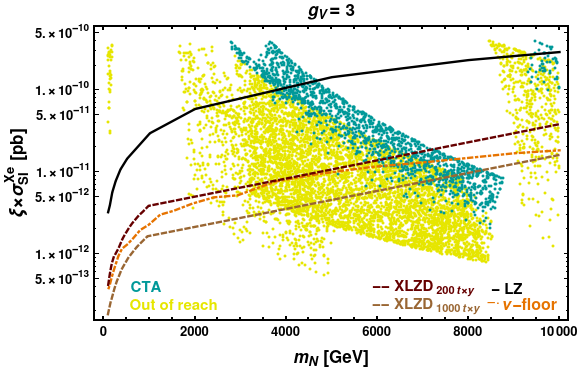}}
  \caption{\small $\xi\times\sigma_{SI}^{Xe}$ w.r.t $m_N$ for $g_{_V}=g,1,2,3$. The thick black line represent the current LZ limit  \cite{LZ:2024zvo}, the brown dashed lines represent the future projection limits of XLZD for exposures $200t\times y$ and $1000t\times y$ \cite{Aalbers:2022dzr}, and the orange dash-dotted one represents the neutrino floor \cite{Billard:2013qya}. The cyan color represent the cases where the points are within  the reach of CTA~\cite{CTA:2020qlo}.}
  \label{fig:SI_DM}
\end{figure*}

\noindent
Before examining the predictions for DM direct detection,  we apply the LEP and LHC constraints on the charged partner, $E^\pm$. For a DM at the electroweak scale the relic density imposes a very small mass splitting which implies a new charged lepton that can easily be produced at colliders. Thus, nearly all the points at the electroweak scale are excluded.
More specifically all points in the region $m_E<m_Z/2$ are excluded by the LEP 1 constraint on the $Z$ boson partial width, while those with $m_E<104$ GeV are excluded by the LEP 2 search for charginos. For  masses around 130 GeV, a few points are compatible with the LHC  searches for long-lived charged particles embedded in {\tt SModelS}. However,  all of these points are excluded by the direct detection limits from LZ~\cite{LZ:2024zvo}. This can be seen in Fig.~\ref{fig:SI_DM} which  shows the prediction for the rescaled SI cross section $\xi\times\sigma_{SI}^{Xe}$ w.r.t $m_N$ for  four benchmark values of $g_{_V}$, together with the current best limit from  LZ~\cite{LZ:2024zvo}. Fig.~\ref{fig:SI_DM} also shows that for DM masses above $1.5$ TeV, a fraction of the points is already excluded by the LZ limits, morever another large fraction is within the reach of future experiments such as XLZD~\cite{Aalbers:2022dzr}. We also note  that an important fraction of the points lie below the neutrino floor. However, we will see in the next section that some will be within reach of future indirect searches.\\

\begin{figure*}[htbp]
  {\includegraphics[width=8cm,height=4cm]{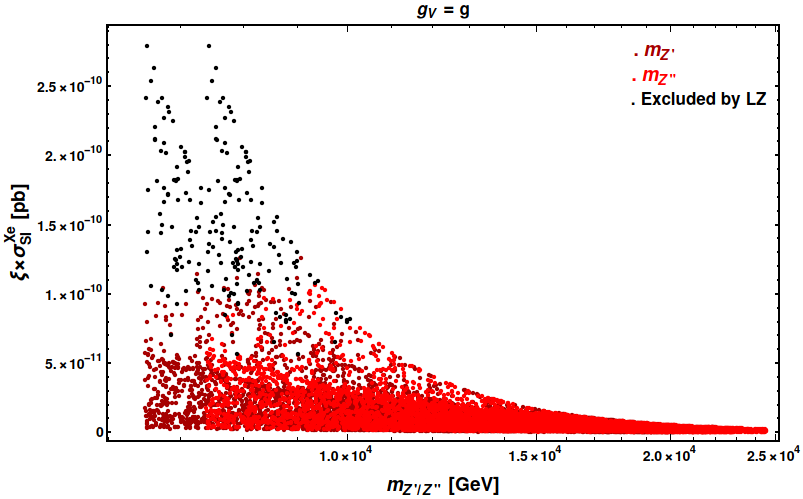}}
  {\includegraphics[width=8cm,height=4cm]{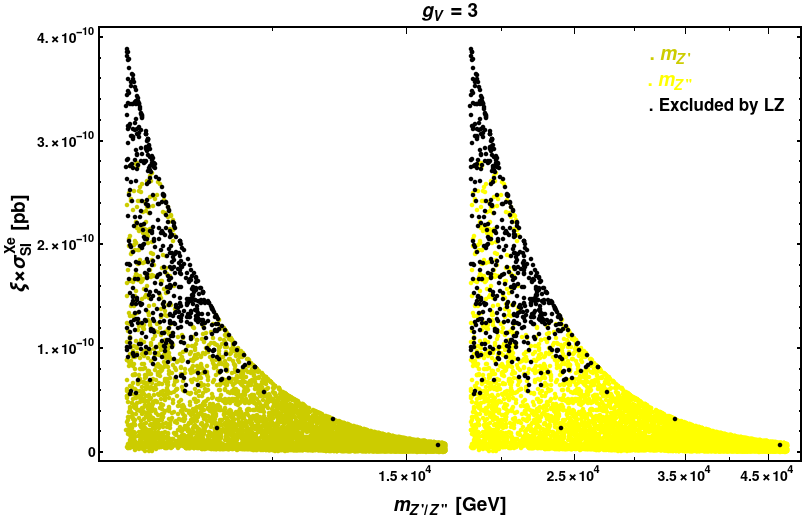}}
  \caption{\small $\xi\times\sigma_{SI}^{Xe}$ variation w.r.t the extra neutral gauge boson masses for two benchmark values of the gauge coupling $g_{_V}=g,3$.}
  \label{fig:SI_MG_Scan}
\end{figure*}

\noindent
The main contribution for the elastic scattering cross section of dark matter on nucleons, comes mainly from diagrams with $Z^{\prime}$ or $Z^{\prime\prime}$, since these contributions go as $1/m_{Z^{\prime}}^4$ or $1/m_{Z^{\prime\prime}}^4$ respectively, the largest cross sections and thus the excluded points by LZ in Fig.~\ref{fig:SI_DM} are  found for the smallest valued of the allowed mass, see Fig.~\ref{fig:SI_MG_Scan}. This figure also illustrates that for large values of $g_{_V}$, the mass splitting between $Z^{\prime}$ and $Z^{\prime\prime}$ increases, thus  one can distinguish the regions where $Z^{\prime}$ or $Z^{\prime\prime}$ give the dominant contribution in Fig.~\ref{fig:SI_DM}. Moreover this large mass splitting entails that for large values of $g_{_V}$ it can become difficult to satisfy the relic density constraint because only one generation of gauge bosons contributes in DM annihilation processes, hence the empty area in the lower panels of Fig.~\ref{fig:SI_DM} around $m_N\approx 8- 9$ TeV.

\newpage
\subsection{Indirect detection bounds}
\noindent
DM can be detected indirectly by observing gamma rays originating from its annihilation in galaxies. As mentioned in section~\ref{DM_int}, the annihilation process could proceed either through s-channel mediated by the vector bosons portal or through t-channels mediated by the VLLs itself (see the Fig~.\ref{FeynD} and Eq.~\ref{final_states}). As discussed before,  the interaction with the $Z$ boson is highly suppressed and can be ignored.   For a TeV scale dark matter mass, the dominant annihilation channels are into quarks, leptons, and pairs of HNs.  These processes proceed via $Z^{\prime}/Z^{\prime\prime}$ exchange in the s-channel. Other possible final states include pairs of neutrinos, HNs with $\nu_i$ , pairs of  neutral or charged gauge bosons, or even $Z/Z^{\prime}$ with $h$. However, all these final states are sub-dominant.\\

\noindent
The photon spectra originating from all final states are computed with {\tt micrOMEGAs} and these spectra are compared with the one  from Dwarf Spheroidal Galaxies observed by Fermi-LAT~\cite{Bonnivard:2015xpq, Alvarez:2020cmw} with the function embedded within {\tt micrOMEGAs}~\cite{Alguero:2023zol}. We find that these data do not allow to constrain the model further, as expected for  TeV scale DM. Here we have not included the points at lower DM masses which are already excluded by DD and collider searches.\\

\begin{figure*}[htbp]
  \centering
  {\includegraphics[width=8cm,height=4cm]{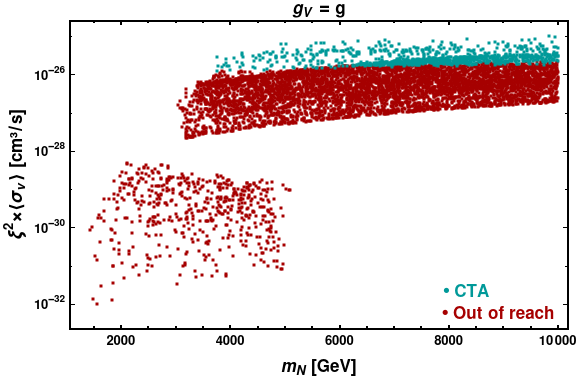}}
   {\includegraphics[width=8cm,height=4cm]{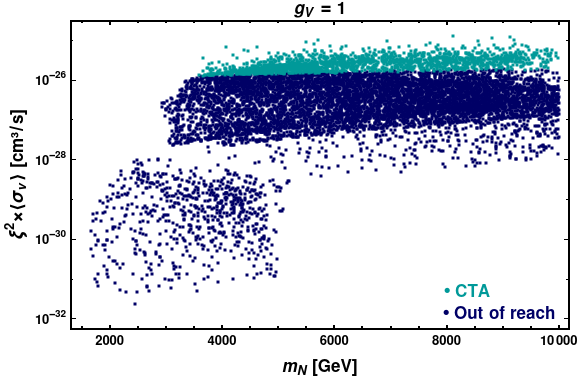}}
    {\includegraphics[width=8cm,height=4cm]{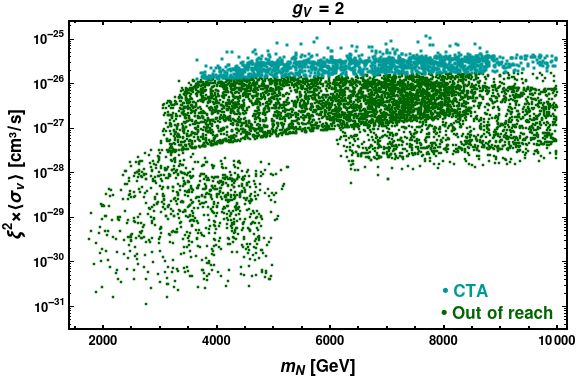}}
     {\includegraphics[width=8cm,height=4cm]{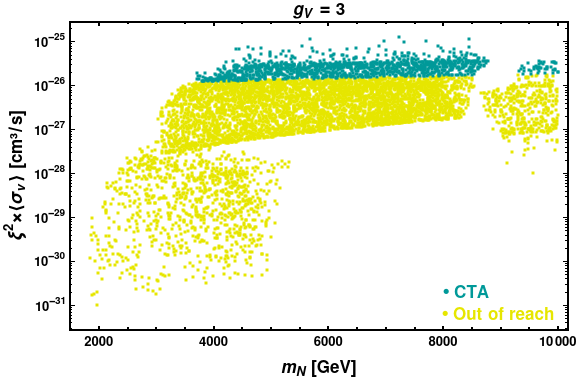}}
  \caption{$\xi^2\times\langle\sigma_{V}\rangle$ w.r.t $m_N$ for several benchmark values of the gauge coupling $g_{_V}$. The cyan color represent scenarios within reach of CTA. The colored points satisfy collider constraints and are beyond the reach of DD.}
    \label{fig:Sv_DM}
\end{figure*}

\noindent
A better probe of TeV scale DM is expected in the future with the CTA telescope.
To determine the potential of CTA to probe the model, we again use {\tt micrOMEGAs} to  compare the total photon spectra corresponding to each scenario with the projections of CTA~\cite{CTA:2020qlo}. Fig.~\ref{fig:Sv_DM} shows the predictions \ for the  rescaled cross section annihilation $\langle\sigma_{V}\rangle$ (i.e. $\xi^2\times\langle\sigma_{V}\rangle$)  as a function of  $m_N$ for the four benchmark values of $g_V$. With the factor $\xi^2$  the cases where DM is under abundant  is taken into account.\\

\begin{figure}[htbp]
  \includegraphics[width=8.0cm,height=5cm]{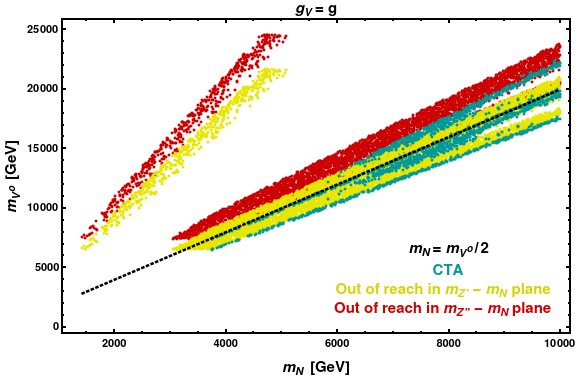}
  \caption{Masses of neutral gauge bosons w.r.t $m_N$ mass for $g_{_V}=g$. The cyan points are within reach of CTA while the red and yellow points are out of reach of DD and ID. All points satisfy collider constraints. The dotted line represent $m_N= m_{Z^{\prime}}/2, m_{Z^{\prime\prime}}/2$.}
  \label{mZpZpp_mN}
\end{figure}

\noindent
In Fig.~\ref{fig:Sv_DM}, the spread of viable points in the range  $m_N\sim1.5-5$ TeV  feature a very suppressed value of $\xi^2\times\langle\sigma_{V}\rangle$. These points correspond to co-annihilation scenarios with s-channel exchange of  $W^{\prime\pm}$ and  a small mass splitting between the component of the VLL doublet. For these points not only  can the relic density  be small ($\xi\ll1$) but also the cross section for pair annihilation of DM can be much suppressed as it does not set the relic density.  For the bulk of the points at the TeV scale, the predictions for  $\xi^2\times\langle\sigma_{V}\rangle$ lie in the range $10^{-28}-10^{-25} {\rm cm}^3/{\rm s}$, thus points in the upper  range falls within the reach of CTA. Note that in this region the relic density is mainly determined by gauge boson exchange,  either a combination of  $Z^{\prime}/Z^{\prime\prime}$  or $W^{\prime\pm},W^{\prime\prime\pm}$ bosons in case of co-annihilation.  For DM pair annihilation mediated through the s-channel exchange of $Z^{\prime}/Z^{\prime\prime}$, the dominant final states are  $q\bar q$, $l^{+}l^{-}$ and  $N_iN_i$. We conclude that a significant fraction of those points  are within  the reach of CTA. The range of masses for $Z^{\prime}/Z^{\prime\prime}$  is illustrated in  Fig.\ref{mZpZpp_mN} and one clearly sees that points with large $\xi^2\times\langle\sigma_{V}\rangle$ within the reach of CTA are close to the region $2m_N\approx m_{Z^\prime}$ or $m_{Z^{\prime\prime}}$. Moreover the distinct region with suppressed $\xi^2\times\langle\sigma_{V}\rangle$ in Fig.\ref{fig:Sv_DM} which do not respect this mass relation can be clearly identified  with both cases of Fig.~\ref{mZpZpp_mN}.

\section{Conclusion}
\label{conc}
\noindent
In this work, we have shown that an extension of  the  LRSM with an  extra $SU(2)_{V}$ factor and  one generation of VLLs can provide a good DM candidate, the neutral component of the VLL doublet.  For stabilizing the DM particle, a discrete parity symmetry that forbids the mixing of VLLs-leptons has been imposed, thus the dark sector particles interact only with vector bosons. DM annihilation and co-annihilation  into fermions proceed mainly through exchange of the vector bosons portal in s-channel otherwise annihilation into final state bosons proceed through VLLs exchange in t-channel. \\

\noindent
After imposing an upper bound on the relic density, we found that DM could be either at the electroweak scale or at the (multi) TeV scale. For the low mass region,  a small mass splitting between the DM particle and its charged partner is required. However, searches for charged leptons at LEP and LHC almost completely exclude this possibility. The only remaining points above $104$ GeV are found to be incompatible with direct detection results of LZ. For DM above the TeV scale, all viable points escape the current limits from dwarf spheroidal galaxies by FermiLAT however the model is partly constrained by direct detection results of LZ and will be further probe by future large detectors. Moreover we have highlighted a complementarity between DD and ID. In particular the future CTA telescope will be able to probe multi-TeV DM that escape multi-ton scale direct detectors such as XLZD and even to probe some points that fall  below the neutrino floor. \\

\noindent  
In this work, we have made simplifying assumptions, for one we only considered the case where  the mass of the HNs  is half that of $m_{W^{\prime}}$ in order to adapt simply the existing LHC limits on new gauge bosons coming from the $W^{\prime\pm}\to N_2\mu^{\pm}$ channel.  We  found that the contribution of final states with  pair of HNs give a non-negligible contribution to DM annihilation  although the annihilation into  quarks and leptons remain dominant.  We checked that removing the HNs from the final state did not affect significantly the photon spectra. We conclude that allowing a larger mass range for the HNs is more likely to impact the DM observables only if it impacts the allowed masses for the new gauge bosons. \\

\noindent 
Finally in our analysis all the masses of extra scalars  were fixed at a high scale, an open question
that we leave for future work is whether taking the scalar sector particles into consideration -- where neutral, charged, and doubly charged scalars could appear in the final states of annihilation and co-annihilation processes -- would impact DM observables. 

\bibliography{biblio}

\end{document}